\documentclass[twocolumn,amsmath,amssymb]{revtex4}
\usepackage{graphicx}
\usepackage{dcolumn}
\usepackage{bm}
\begin{document}

\title{Fractal Butterflies of Dirac Fermions in Monolayer and Bilayer graphene}
\author{Tapash Chakraborty$^\ddag$}
\affiliation{Department of Physics and Astronomy,
University of Manitoba, Winnipeg, Canada R3T 2N2}
\author{Vadym M. Apalkov}
\affiliation{Department of Physics and Astronomy, Georgia State University,
Atlanta, Georgia 30303, USA}

\date{\today}
\begin{abstract}
We present an overview of the theoretical understanding of Hofstadter butterflies in monolayer and
bilayer graphene. After a brief introduction on the past work in conventional semiconductor
systems, we discuss the novel electronic properties of monolayer and bilayer graphene that helped 
to detect experimentally the fractal nature of the energy spectrum. We have discussed the theoretical
background on the Moir\'e pattern in graphene. This pattern was crucial in determining the butterfly
structure. We have also touched upon the role of electron-electron interaction in the butterfly 
pattern in graphene. We conclude by discussing the future prospects of butterfly search, especially 
for interacting Dirac fermions.
\end{abstract}
\maketitle

\section{Introduction}

The dynamics of an electron in a periodic potential subjected to a perpendicular magnetic field has 
remained an interesting problem for more than half a century \cite{review_2000,harper,langbein,hofstadter}. 
Within the nearest neighbor tight-binding description of the periodic potential
the energy spectrum of an electron is described by the Harper equation \cite{harper}.
Numerical solution of this equation \cite{hofstadter} shows that the applied magnetic 
field splits the Bloch bands into subbands and gaps. The resulting energy spectrum, when plotted 
as a function of the magnetic flux per lattice cell, reveals a fractal pattern (a self-similar 
pattern that repeats at every scale) \cite{mandelbrot} that is known in the literature as Hofstadter's 
butterfly (due to the pattern resembling the butterflies). This is the first example of the fractal 
pattern realized in the energy spectra of a physical system. 

A few experimental efforts to detect the butterflies have been reported in the literature. The 
earlier ones involved artificial lateral superlattices on semiconductor nanostructures 
\cite{geisler_04_1,geisler_04_2,albrecht_01,albrecht_02,ensslin_96_1,ensslin_96_2}, more precisely 
the antidot lattice structures with periods of $\sim$100 nm. The large period (as opposed to those 
in natural crystals) of the artificial superlattices helps to keep the 
magnetic field in a reasonable range of values to observe the fractal pattern. Measurements 
of quantized Hall conductance in such a structure indicated, albeit indirectly, the complex pattern
of gaps that were expected in the butterfly spectrum. Hofstadter butterfly patterns were also 
predicted to occur in other totally unrelated systems, such as, propagation of microwaves 
through a waveguide with a periodic array of scatterers \cite{microwave} or more recently, with 
ultracold atoms in optical lattices \cite{optical_lattice_1,optical_lattice_2}.

Graphene, the single layer of carbon atoms, arranged in a hexagonal lattice and contains an
wealth of unusual electronic properties \cite{graphene_book,abergeletal,chapter,xu_review} has turned 
out to be the ideal system in the quest of fractal butterflies. The Dirac fermions in monolayer and bilayer 
graphene \cite{chapter} are the most promising objects thus far, where the signature of the recursive 
pattern of the Hofstadter butterfly has been unambiguously reported \cite{dean_13,hunt_13,geim_13}. 
Here the periodic lattice with a period of $\sim$ 10 nm was created by the Moir\'e pattern that appears 
when graphene is placed on a plane of hexagonal boron nitride (h-BN) with a twist \cite{hbn,moire_1,moire_2}. 
Being ultraflat and free of charged impurities, h-BN has been the best substrate for graphene
having high-mobility charged fermions \cite{hbn}. Some theoretical studies have been reported earlier
in the literature on the butterfly pattern in monolayer \cite{rhim} and bilayer graphene \cite{nemec}. 

The paper is organized as follows. In Sect. II, we briefly describe the background materials leading
to the Hofstadter butterfly. The situation in conventional semiconductor systems is presented in
Sect. III. Sect. IV deals with the theories of the butterfly pattern in monolayer graphene, while the 
theoretical intricacies in bilayer graphene are presented in Sect. V. The case of the many-electron
system, in particular the influence of the electron-electron interaction on the Hofstadter butterfly
pattern is described in Sect. VI. The concluding remarks are to be found in Sect. VII.

\section{Electrons in a periodic potential and an external magnetic field: Hofstadter butterfly} 

The dynamics of a two-dimensional (2D) electron in a periodic potential is described by the Hamiltonian 
\begin{equation}
{\cal H} = {\cal H}^{}_0 (p^{}_x,p^{}_y) + V(x,y),
\label{Hgeneral}
\end{equation}
which consists of the kinetic energy term ${\cal H}^{}_0 (p^{}_x,p^{}_y)$ and the periodic potential 
$V(x,y)$.  The most important characteristics of this periodic potential, which determines the dynamics 
of an electron in a magnetic field is the area $S^{}_0$ of the unit cell of the periodic structure of 
$V(x,y)$. For a structure of a simple square lattice type, which is characterized by the lattice 
constant $a^{}_0$, the area of a unit cell is $S^{}_0 = a_0^2$. The magnetic field $B$ is introduced
in the Hamiltonian (\ref{Hgeneral}) via the Peierls substitution, which replaces the momentum 
$(p^{}_x,p^{}_y)$ by the generalized expression $(p^{}_x-eA^{}_x/c,p^{}_y-eA^{}_y/c)$. Here 
$(A^{}_x,A^{}_y)$ is the vector potential. We choose the vector potential in the Landau gauge  
$\vec{A} = (0,Bx)$. The corresponding Hamiltonian then becomes 
\begin{equation}
{\cal H} = {\cal H}^{}_0 (p^{}_x-exB) + V(x,y).
\label{Hmagnet}
\end{equation}

\begin{figure}
\begin{center}\includegraphics[width=8cm]{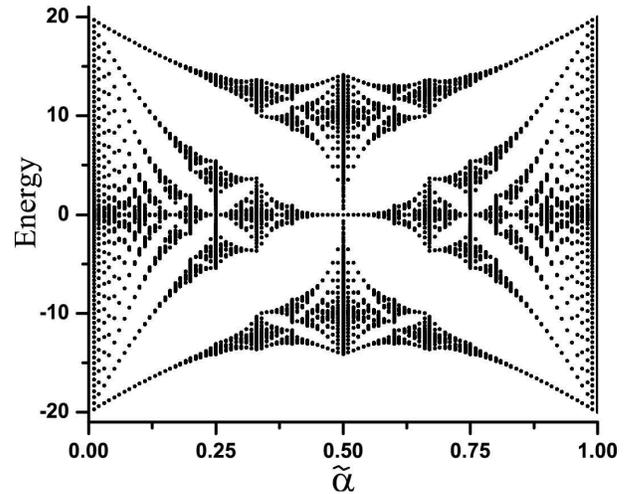}\end{center}
\caption{Energy spectra (Hofstadter butterfly) of the Harper equation (\ref{Harper}). 
Parameter $\tilde{\alpha}$ is the magnetic flux per unit cell in units of flux quantum.  
}
\label{butterfly}
\end{figure}

The energy spectra of the Hamiltonian (\ref{Hmagnet}) as a function of the magnetic field has the
unique fractal structure. Such a structure has a more clear description in the two limiting cases of weak 
and strong magnetic field. In the case of the weak magnetic field, first, the periodic potential results 
in the formation of the Bloch bands and then the external magnetic field splits each Bloch band of 
the periodic potential into minibands of the Landau level (LL) type. In a weak magnetic field, the coupling 
of different bands can be disregarded. The corresponding Schr\"odinger equation, which determines the 
energy spectrum of the system, has a simple form in the tight-binding approximation 
for the periodic potential, for which the energy dispersion within a single band is 
\begin{equation}
E(p^{}_x,p^{}_y) = 2\Delta^{}_0\left( \cos (p^{}_xa^{}_0/\hbar) + \cos (p^{}_ya^{}_0/\hbar) \right) ,
\label{EsingleBand}
\end{equation}
where a simple square lattice structure with lattice constant $a^{}_0$ was assumed. In an external magnetic field, 
the wave function which is defined at the lattice points $(ma^{}_0,na^{}_0)$, has the form $\Psi(ma,na) = 
e^{ik^{}_y n} \psi^{}_m$. The corresponding Schr\"odinger equation reduces to a one-dimensional equation -- the so
called Harper equation \cite{harper}
\begin{equation}
\psi^{}_{m+1} + \psi^{}_{m-1} + 2 \cos \left( 2\pi m \tilde{\alpha} - k^{}_y \right)\psi^{}_{m} = 
\varepsilon \psi^{}_m,
\label{Harper}
\end{equation}
where $\varepsilon = E/\Delta^{}_0$ and $\tilde{\alpha} = \Phi /\Phi^{}_0$. Here $\Phi = BS^{}_0 = 
Ba_0^2$ is the magnetic flux through a unit cell and $\Phi^{}_0 = hc/e$ is the magnetic flux quantum. 
Therefore, the dimensionless parameter $\tilde{\alpha}$ is the magnetic flux through a unit cell measured in 
units of the flux quantum. The energy spectra, determined by the Harper equation (\ref{Harper}), is a 
periodic function of the parameter $\tilde{\alpha}$ with period 1. Hence it is enough to consider only 
the values of $\tilde{\alpha}$ within the range $0<\tilde{\alpha}<1$. The remarkable property of the 
Harper equation (\ref{Harper}) is that although the corresponding Hamiltonian is an analytical function 
of $\tilde{\alpha}$, the energy spectrum [Eq. (\ref{Harper})] is very sensitive to the value of 
$\tilde{\alpha}$. At rational values of $\tilde{\alpha} = p/q$ the energy spectrum has $q$ bands 
separated by $q-1$ gaps, where each band is $p$ fold degenerate. As a function of $\tilde{\alpha}$ the 
energy spectrum [Eq.\ (\ref{Harper})] has a fractal structure that is known as the Hofstadter butterfly 
\cite{hofstadter}. This structure is shown in Fig.\ \ref{butterfly}. The thermodynamic potential 
$\Omega^{}_b(T,\mu, \tilde{\alpha})$, corresponding of the system described by the Harper equation 
(\ref{Harper}), satisfies the following symmetry property \cite{avron}
\begin{equation}
\Omega^{}_b(T,\mu, \tilde{\alpha}) = \Omega^{}_b(T,\mu, -\tilde{\alpha})= \Omega^{}_b(T,\mu, 
\tilde{\alpha}+1), 
\end{equation}
which means that the thermodynamic properties of the system are determined by $0<\tilde{\alpha} <1/2$ 
and $\mu <0$. Here $\mu$ is the chemical potential. 

In a strong magnetic field the energy spectra of the system also show the Hofstadter butterfly 
fractal structure. Now the periodic potential should be considered as a weak perturbation, 
which results in a splitting of the corresponding LLs, formed by the strong magnetic field. 
For a weak periodic potential the inter LL coupling can be disregarded. Then the splitting of 
a given LL is described by the same Harper-type equation,
\begin{equation}
\psi^{}_{m+1} + \psi^{}_{m-1} + 2 \cos \left( 2\pi m \alpha - k^{}_y \right)\psi^{}_{m} = \varepsilon
\psi^{}_m,
\label{Harper2}
\end{equation}
but now the parameter, which determine the fractal structure of the energy spectrum, is $\alpha = 
1/\tilde{\alpha}= \Phi^{}_0/\Phi$ - inverse magnetic flux though a unit cell in units of the flux 
quantum. Therefore, for $0<\alpha<1$ the energy spectrum has a structure similar to the 
one shown in Fig. \ref{butterfly}. The Hofstadter butterfly energy spectra is realized either 
as a splitting of the Bloch band by a weak magnetic field or as a splitting of a LL by the weak 
periodic potential. The thermodynamics properties of these two systems are related by the duality 
transformation 
\cite{avron}
\begin{equation}
\Omega^{}_L(T,\mu, \alpha ) = \alpha  \Omega^{}_b(T,\mu, \tilde{\alpha}),
\end{equation}
where $\Omega^{}_L(T,\mu, \alpha )$ is the thermodynamic potential within a single LL and weak periodic 
potential. 

For intermediate values of the magnetic field, the mixing of the LL by the periodic potential or 
the mixing of Bloch bands by the magnetic field becomes strong. This will modify the universality of 
the butterfly structure and add some system-dependent features. In the following sections we consider the 
limits of high and intermediate magnetic fields for conventional semiconductor systems and the monolayer 
and bilayer graphene. 

\begin{figure}
\begin{center}\includegraphics[width=7cm]{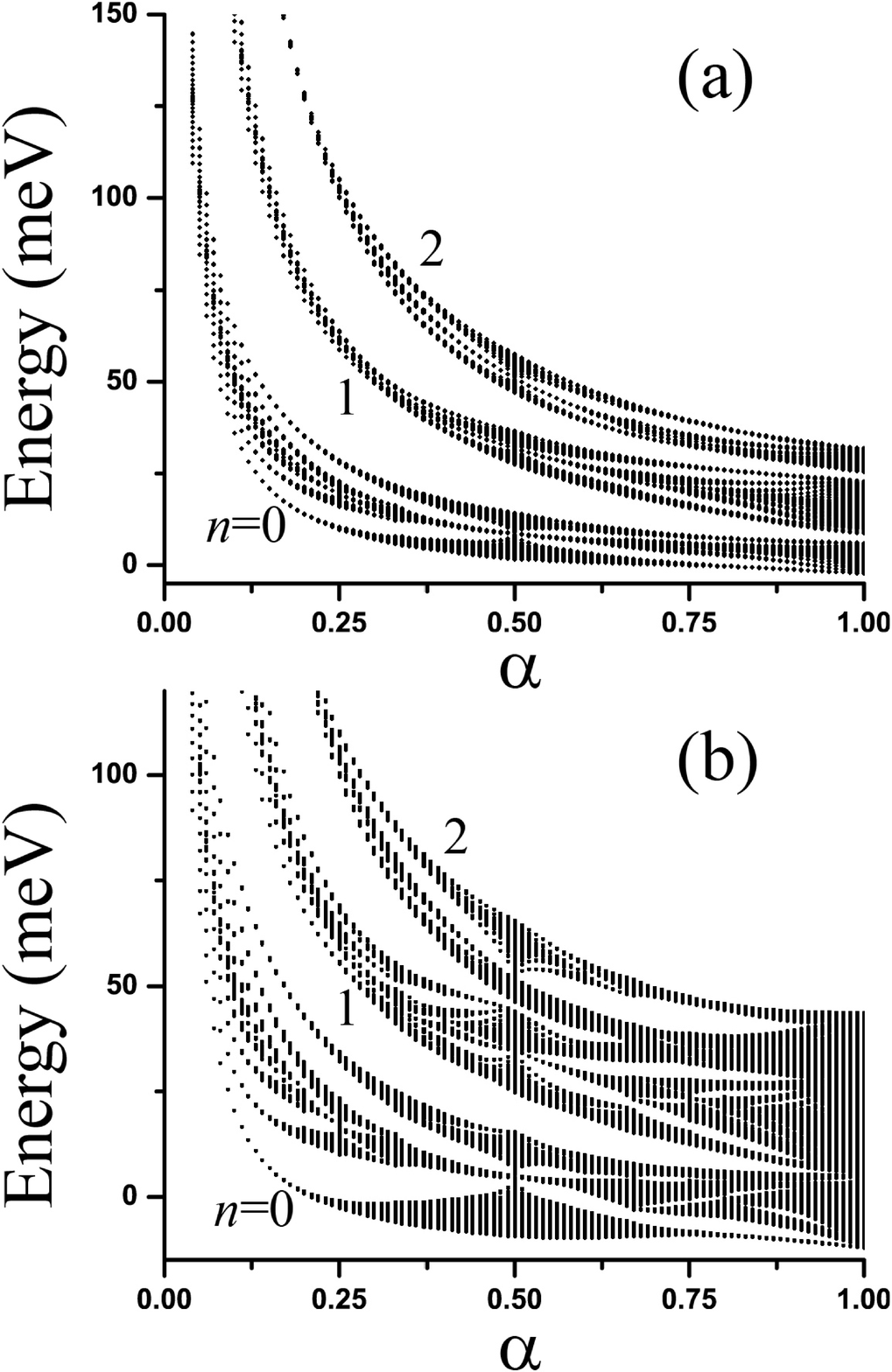}\end{center}
\caption{Single-electron energy spectra of conventional semiconductor systems with parabolic dispersion 
relation.  The period of the potential is $a^{}_0 = 20$ nm and its amplitude is (a) $V^{}_0 = 10$ meV 
and (b) $V^{}_0 = 20$ meV. The energy spectra are shown as a function of the parameter $\alpha = 
\Phi^{}_0 /\Phi$. The numbers indicate the LL index $n$. 
}
\label{Fig_conventional}
\end{figure}

\section{Conventional semiconductor systems: strong field limit}

For strong and intermediate magnetic fields, the periodic potential is considered as a perturbation, 
which can modify and mix the states of the zero-order Hamiltonian, consisting of the kinetic part only 
${\cal H}^{}_0 (p^{}_x-eA^{}_x/c,p^{}_y-eA^{}_y/c)$. For conventional semiconductor systems 
the zero-order Hamiltonian is described by the parabolic dispersion relation, $p^2/2m$. The 
transverse magnetic field results in Landau quantization where the LLs are 
characterized by the Landau level index $n=0,1,2,\ldots$ with energies $E^{}_n = (n+1/2)\hbar
\omega^{}_{c,B}$. Here $\omega^{}_{c,B}=eB/mc$ is the cyclotron frequency. The corresponding Landau 
wave functions $\phi^{}_{n,k}$ have the form
\begin{equation}
\phi^{}_{n,k} (x,y) = \frac{e^{i k y}}{\sqrt{L}} 
\frac{e^{-(x-x^{}_k )^2/2\ell_0^2}}{\sqrt{ \pi^{1/2} \ell^{}_0 2^n n!}}
 H^{}_n (x-x^{}_k),
\label{PhiK}
\end{equation}
where $L$ is the length of a sample in the $y$ direction, $k$ is the $y$ component of the electron 
wave vector, $\ell^{}_0 = \sqrt{c\hbar/eB}$ is the magnetic length, $x^{}_k=k\ell^{2}_0$, and 
$H^{}_n(x)$ are the Hermite polynomials. 

We consider a system in a periodic external potential that has the form  
\begin{equation}
V(x,y) = V^{}_0 \left[\cos(q^{}_x x)+\cos(q^{}_y y)\right],
\label{Vxy}
\end{equation}
where $V^{}_0$ is the amplitude of the periodic potential, $q^{}_x = q^{}_y = 
q^{}_0 = 2\pi/a^{}_0$, and $a^{}_0$ is a period of the external potential  $V(x,y)$.
The periodic potential mixes the electron states $\phi^{}_{n,k}$ within a single LL, i.e., states with 
the same value of LL index $n$ and different values of $k$, and also mixtures the states of different 
LLs with different indices $n$. The strength of the mixing is determined by the matrix elements of the 
periodic potential $V(x,y)$ between the LL states $\phi^{}_{n,k}$.

The matrix elements of the periodic potential $V(x,y)$ in the basis 
$\phi^{}_{n,k} (x,y)$ are
\begin{eqnarray}
& & \left\langle \phi^{}_{n^{\prime },k^{\prime }} \right|\cos(q^{}_0 y)\left|\phi^{}_{n,k}\right
\rangle = \nonumber \\  
& & \frac{i^{n-n^{\prime}}}{2}  
\left\{\delta^{}_{k^{\prime}, k+q^{}_0} + (-1)^{n-n^{\prime}} \delta^{}_{k^{\prime}, k-q^{}_0} \right\}  
M^{}_{n^{\prime },n}
\label{Vy0}
\end{eqnarray}
and 
\begin{equation}
\left\langle \phi^{}_{n^{\prime },k^{\prime }} \right|\cos(q^{}_0 x) \left| \phi^{}_{n,k} \right
\rangle = \frac{\delta^{}_{k^{\prime}, k}}{2} e^{-i q^{}_0 k\ell_0^2} \left[1+(-1)^{n-n^{\prime}}
\right] M^{}_{n^{\prime},n}.
\label{Vx0}
\end{equation}
Here 
\begin{equation}
M^{}_{n^{\prime},n} = \left(\frac{m!}{M!}\right)^{1/2} e^{-Q/2} Q^{|n^{\prime}
-n|/2} L_m^{|n^{\prime }-n|} (Q),
\end{equation}
$Q = q_0^2 \ell_0^2/2$, $m=\min (n^{\prime},n)$, $M=\max (n^{\prime},n)$.

The matrix elements (\ref{Vy0}) and (\ref{Vx0}) determine the mixture of the LL states introduced by 
the periodic potential. While the component of the potential periodic in the $x$ direction [Eq. (\ref{Vx0})] 
couples only the states with the same value of the wave vector $k$, the component periodic in the $y$ 
direction couples the states with the wave vectors separated by $q^{}_0$. Within a single LL the potential 
periodic in the $x$ direction modifies the energy of each Landau state. As a result the energy of the Landau 
state within a given LL becomes a periodic function of $q^{}_0kl_0^2$. Additional coupling of the states 
separated by $q^{}_0$, which is determined by Eq.\ (\ref{Vy0}), results in the formation of the band 
structure when $q_0^2l_0^2$ becomes a rational fraction of $2\pi$, which is exactly the condition that the 
parameter $\alpha $ is rational. It follows from Eqs.\ (\ref{Vy0})-(\ref{Vx0}) that for a given LL with 
index $n$ the effective amplitude of the periodic potential acquires an additional factor and becomes 
$\propto V^{}_0 M^{}_{n,n} \propto L^{}_n(q_0^2 \ell_0^2/2) = L^{}_n (\pi \alpha)$. These renormalized 
amplitudes determine the width of the corresponding bands. At values of $\alpha$ where $L^{}_n (\pi 
\alpha) = 0$, all bands have zero width which correspond to the flatband condition 
\cite{geisler_04_1,geisler_04_2}.

In general, the expressions for the matrix elements (\ref{Vx0}) and (\ref{Vy0}) can be used to find the 
energy spectra of any finite number of LLs, taking into account the coupling of different LLs 
introduced by the periodic potential. For a given value of $k$ within the interval $0<k<q^{}_0$, a finite 
set of basis wave functions $\phi^{}_{n,k}$, $\phi^{}_{n,k+q^{}_0}$, $\phi^{}_{n,k+2q^{}_0}$,\ldots, 
$\phi^{}_{n,k+N^{}_xq^{}_0}$ is considered. Here $n=0,\ldots, N^{}_L$, $N^{}_L$ is the number of LLs,
and $N^{}_x$ determines the size of the system in the $x$ direction: $L^{}_x = N^{}_x q^{}_0 \ell_0^2$. 
The matrix elements (\ref{Vy0}) and (\ref{Vx0}) determine the coupling of the states within this 
truncated basis and finally determines the corresponding Hamiltonian matrix. The diagonalization of the
matrix provides the energy spectrum for a given value of $k$. The spectra are calculated for a finite number 
$N^{}_y$ of $k$ points, where $N^{}_y$ determines the size of the system in the $y$ direction: 
$L^{}_y = 2\pi N^{}_y /q^{}_0$. 

Following this procedure the energy spectra of the conventional system with parabolic dispersion relation 
were evaluated for $N^{}_L = 2$ LLs. The results are shown in Fig.\ \ref{Fig_conventional} for the 
period of the potential $a^{}_0 = 20 $ nm. The results clearly show that although for the potential 
amplitude $V^{}_0 = 10 $ meV the mixing of LLs is relatively weak, for a higher amplitude  $V^{}_0 = 20$ 
meV the mixing becomes strong especially for $\alpha$ close to 1, i.e., in weak magnetic fields. The 
butterfly structure is no longer described by the simple Harper equation. In Ref.\ \cite{butterfly_exact} 
a detailed analysis of the Hofstadter butterfly spectrum was done for strong and intermediate periodic 
potential strength. The magnetic field splits the Bloch bands and introduces coupling of the states of 
different Bloch bands.  

\section{Monolayer graphene}

\subsection{Square lattice periodic structure}

The unique feature of graphene is a relativistic-like low-energy dispersion relation 
\cite{graphene_book,abergeletal}, corresponding to the Dirac fermions \cite{chapter},
which results in several unique features in Landau quantization and in the structure of the LLs. The 
LLs in graphene have two-fold valley degeneracy corresponding to two valleys $K$ and $K^{\prime}$. 
The degeneracy cannot be lifted by periodic potential with typical long periods, $a^{}_0 > 10 $ nm. 
In this case the Hofstadter butterfly pattern realized in graphene have two-fold valley degeneracy 
and it is enough to consider only the states of one valley, e.g., valley $K$. The corresponding 
Hamiltonian ${\cal H}^{}_0$ is written \cite{graphene_book,abergeletal} in the matrix form 
\begin{equation}
{\cal H}^{}_0 = v^{}_F \left( 
\begin{array}{cc}
    0 & \pi^{}_x - i \pi^{}_y   \\
    \pi^{}_x + i \pi^{}_y & 0 
\end{array} 
\right) ,
\label{HB}
\end{equation}
where $\vec{\pi } = \vec{p} + e\vec{A}/c$, $\vec{p}$ is the electron momentum and $v^{}_F \approx 10^6$ 
m/s is the Fermi velocity. 

The LLs in graphene, which are determined by the Hamiltonian (\ref{HB}), are specified by the Landau 
index $n=0, \pm 1, \pm 2, \ldots$, where the positive and negative values correspond to the conduction and 
valence band levels, respectively. The energy of the LL with index $n$ is \cite{graphene_book,abergeletal}
\begin{equation}
E^{(gr)}_n = s^{}_n \hbar \omega^{}_{gr,B} \sqrt{|n|},
\end{equation}
where $\omega^{}_{gr,B} = v^{}_F/\ell^{}_0$ is the cyclotron frequency in graphene; $s^{}_n=1$ for
$n>1$, $s^{}_n =0$ 
for $n=0$, and $s^{}_n =-1$ for $n<1$. 

\begin{figure}
\begin{center}\includegraphics[width=7cm]{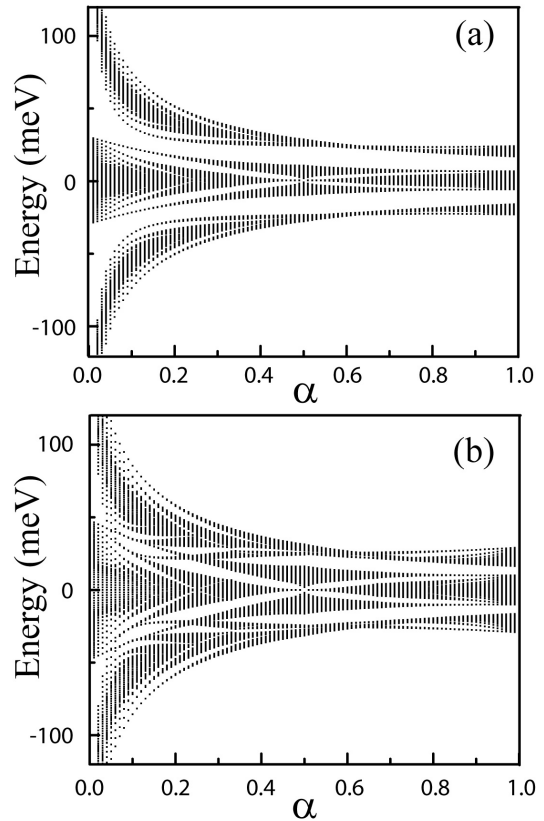}\end{center}
\caption{Single-electron energy spectra of graphene monolayer in a periodic potential and an external
magnetic field. The period of the potential is $a^{}_0 = 20$ nm and its amplitude is (a) $V^{}_0 =
50$ meV and (b) $V^{}_0 = 100$ meV. The energy spectra are shown as a function of the parameter
$\alpha = \Phi^{}_0 /\Phi$.
}
\label{Fig_graphene}
\end{figure}

The eigenfunctions of the Hamiltonian (\ref{HB}), corresponding to the LL with index $n$, are given by 
\begin{equation}
\Psi^{}_{n,k} = C^{}_n
\left( \begin{array}{c}
 s^{}_n i^{|n|-1} \phi^{}_{|n|-1,k} \\
    i^{|n|} \phi ^{}_{|n|,k}
\end{array}  
 \right),
\label{f1}
\end{equation}
where $C^{}_n = 1 $ for $n=0$ and $C^{}_n = 1/\sqrt{2}$ for $n\neq 0$. Here $\phi^{}_{n,k}$ is the 
Landau wave function introduced by Eq.\ (\ref{PhiK}) for an electron with parabolic dispersion relation. 
The graphene monolayer is then placed in a weak periodic potential $V(x,y)$, which is given by Eq.\ (\ref{Vxy}).
This potential introduces coupling of LLs in graphene. The corresponding matrix elements of the periodic 
potential are
\begin{eqnarray}
& & \left\langle n^{\prime }k^{\prime}\right|\cos(q^{}_0 y)\left| n k\right
\rangle = \nonumber \\  
& & \frac{i^{n-n^{\prime}}}{2} C^{}_n C^{}_{n^{\prime}} 
\left\{\delta^{}_{k^{\prime}, k+q^{}_0} 
  +  (-1)^{n-n^{\prime }} \delta^{}_{k^{\prime}, k-q^{}_0} 
  \right\}  \times    \nonumber \\
 & &  \left[ s^{}_n s^{}_{n^{\prime}}  M^{}_{|n^{\prime }|-1,|n|-1} + 
   M^{}_{|n^{\prime }|,|n|} \right] 
\end{eqnarray}
and 
\begin{eqnarray}
& & \left\langle n^{\prime}k^{\prime}\right|\cos(q^{}_0 x) \left| n k\right
\rangle = \frac{\delta^{}_{k^{\prime}, k}}{2} C^{}_n C^{}_{n^{\prime}} 
e^{-i q^{}_0 k\ell_0^2} \\
& &\times \left[1+(-1)^{n-n^{\prime}}\right] 
\left[s^{}_n s^{}_{n^{\prime}} M^{}_{|n^{\prime}|-1,|n|-1} + 
 M^{}_{|n^{\prime}|,|n|} \right]. \nonumber
\end{eqnarray}

For a given LL with index $n$, the periodic potential is determined by the effective value 
\begin{eqnarray}
& & V^{}_0 \left[s_n^2 M^{}_{|n|-1,|n|-1} +   M^{}_{|n|,|n|} \right] \nonumber \\
& & \propto V^{}_0 \left[s_n^2 L^{}_{|n|-1} (\pi \alpha) +  L^{}_{|n|} (\pi \alpha)  \right].
\end{eqnarray}
The flatbands in graphene are therefore realized at points where $s_n^2 L^{}_{|n|-1} (\pi\alpha) + L^{}_{|n|} 
(\pi \alpha)$ is 0. For $n=0$, i.e., $s^{}_0 = 0$, this is exactly the same condition as in conventional 
system, but for other LLs the condition of flatbands becomes $ L^{}_{|n|-1} (\pi \alpha) +  L^{}_{|n|} 
(\pi \alpha) =0$. 

In Fig.\ \ref{Fig_graphene} the Hofstadter butterfly energy spectra is shown for a graphene monolayer, 
taking into account three LLs with $n=-1$, 0, and 1. The main difference between the conventional
systems and graphene is the broadening of the energy structure within a single LL. For conventional 
system [Fig.\ \ref{Fig_conventional}], the width of the energy spectra for the $n=1$ LL is small for 
small values of $\alpha$ and large for large $\alpha$. In graphene the behavior is different: the 
broadening of the $n=1$ LL is large for small values of $\alpha$ and small for intermediate and large 
values of $\alpha$. Another specific feature of the energy spectra of graphene is that the mixing of 
the LLs, introduced by the periodic potential, is visible for much large values of the amplitude of 
the potential, $V^{}_0 \approx 100 $ meV compared to $V^{}_0 \approx 20$ meV in conventional systems 
[Fig.\ \ref{Fig_conventional}(b)].

\subsection{Moir\'e structure}

\begin{figure}
\begin{center}\includegraphics[width=7cm]{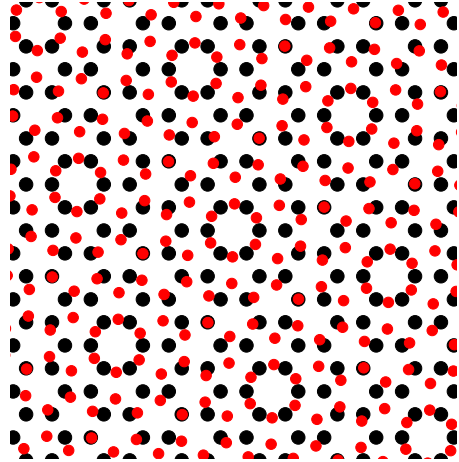}\end{center}
\caption{Moir\'e pattern in two hexagonal lattices with rotational misalignment. The two lattices,
which correspond to two layers are shown by red and black dots, respectively.
}
\label{Fig_moire}
\end{figure}

With the system of graphene one has the unique possibility to generate in the Hamiltonian 
a periodical perturbation (periodic potential) based on the intrinsic structure of graphene-based
systems. Such a periodic structure is based on the Moir\'e pattern which appears between two similar regular 
structures overlaid at an angle. In graphene, the Moir\'e pattern is realized in (i) twisted bilayer 
graphene \cite{stacking_geometry_06,bilayer_with_twist_07,misoriented,multi_layer_08,quantum_interference_08,
Van_hove_twisted_10,Mele_moire_2010,trasport_twisted_graphene,bistritzer,luican} which consists of two 
monolayers with relative small rotation angle between the layers; (ii) graphene monolayer on hexagonal 
boron nitride substrate with rotational misalignment between the graphene monolayer and the h-BN 
\cite{hbn,zero_energy_mode,graphene_moire_parten,dean_13,hunt_13,geim_13}. Realization of the Moir\'e 
pattern in two hexagonal lattices (layers) is shown in Fig.\ \ref{Fig_moire}. That pattern introduces 
a large-scale periodicity in the Hamiltonian of the systems, which, in a magnetic field, results in
the Hofstadter butterfly spectra. 

For twisted bilayer graphene the periodical modulation of the Hamiltonian is introduced through the interlayer 
hopping coupling which capture the periodic structure of the Moir\'e pattern. The interlayer coupling 
matrix is \cite{bistritzer}
\begin{equation}
T(\vec{r}) = w\sum_j e^{-i\vec{q}_j \vec{r}} T^{}_j,
\end{equation}
where $j=1,2,3$ and matrices $T^{}_j$ have the form 
\begin{equation}
T^{}_1 = \left( \begin{array}{cc}
1 & 1 \\
1 & 1
\end{array} 
\right),\,
T^{}_2 = \left( \begin{array}{cc}
e^{-i\psi }  & 1 \\
e^{i\psi } & e^{-i\psi }
\end{array}
\right), \,
T^{}_3 = \left( \begin{array}{cc}
e^{i\psi }  & 1 \\
e^{-i\psi } & e^{i\psi }
\end{array}
\right).
\end{equation}
Here $\psi = 2\pi/3$, $\vec{q}^{}_1 = k^{}_D \theta (0,-1)$,  $\vec{q}^{}_2 = k^{}_D \theta 
(\sqrt{3}/2,1/2)$, $\vec{q}^{}_3 = k^{}_D \theta (-\sqrt{3}/2,1/2)$, $\theta$ is the twist angle, 
$k^{}_D$ is the Dirac momentum, and $w$ is the hopping energy. The interlayer coupling has a matrix form, 
where the two components of the matrix correspond to two layers of graphene bilayer. 

The Moir\'e periodicity in the matrix $T(\vec{r})$ results in the formation of the Hofstadter butterfly pattern, 
which was studied in Refs. \cite{bistritzer} as a splitting of the Landau levels due to the weak periodical 
modulation of $T(\vec{r})$. Since the area of the Moir\'e  units cell is $\propto 1/\theta^2$, to observe the 
Hofstadter butterfly pattern for experimentally realized magnetic fields the twist angle should be
small, $\theta \lesssim 5^0$. 

Just as for the twisted bilayer graphene, the periodical perturbation in the Hamiltonian of monolayer graphene placed 
on a h-BN substrate is introduced through the periodical modulation of the interlayer coupling. The difference from 
the bilayer graphene case is that there is a small $\approx 1.8$\% lattice mismatch between graphene and the BN. As 
a result, the interlayer coupling is determined by both the lattice mismatch and rotational misalignment by an 
angle $\theta$. Then the corresponding superlattice period $a^{}_0$ depends both on the twist angle and
the lattice mismatch. Even in the case of perfect alignment, i.e., for the zero twist angle, the superlattice period 
is $a^{}_0 \approx 13 $ nm. This value introduces upper limits on the superlattice period. This is different 
from twisted graphene bilayer, for which there is no superllatice for perfect alignment of the layers and there 
is no constraint on the values of $a^{}_0$. Another specific feature of graphene monolayer on the h-BN substrate 
is an asymmetry term in the effective Hamiltonian of graphene, which is due to different couplings of the B and 
N atoms to the graphene layer. 
 
\begin{figure}
\begin{center}\includegraphics[width=7cm]{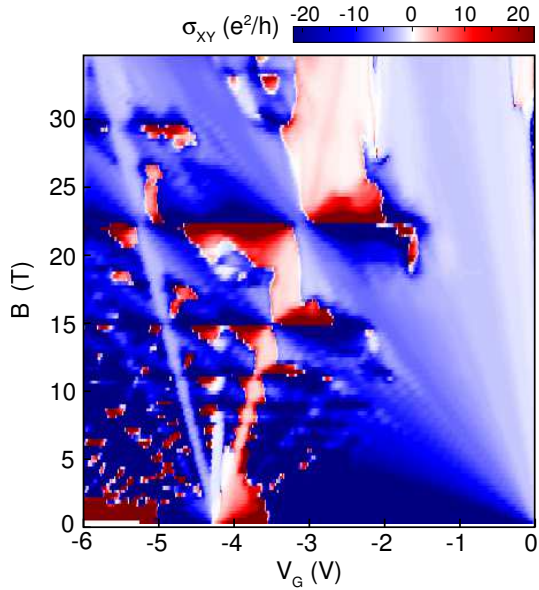}\end{center}
\caption{Experimental results for the Hall conductance probe of minigap opening within a Landau level in 
graphene, depicting the self-similarity pattern. (Courtesy of P. Kim and C. Dean). 
}
\label{dean}
\end{figure}

The periodic perturbation of the graphene Hamiltonian on the h-BN substrate, i.e., the graphene superlattice, 
results in the formation of multiple Moir\'e minibands and generation of secondary Dirac points
\cite{zero_energy_mode,graphene_moire_parten,Dirac_edges_graphene_moire} near the edges of the superlattice 
Brillouin zone. These points are characterized by the wave vector $G=4\pi/\sqrt{3} a^{}_0$. The energy 
corresponding to this vector is $E^{}_G = \hbar v^{}_F G/2$. To observe these secondary Dirac points the 
graphene should be doped upto energy $E^{}_G$. Since the period of the Moir\'e superlattice is determined by the 
twist angle, the doping requirement introduces a constraint on the values of the twist angle, which should be 
less than $1^0$ \cite{geim_13}. The formation of the fractal Hofstadter butterfly pattern in graphene on
the h-BN substrate was studied theoretically in Ref.\ \cite{Dirac_edges_graphene_moire} and was later observed 
experimentally in Refs.\ \cite{hunt_13,dean_13,geim_13}. This butterfly pattern was realized as splitting of 
the Moir\'e minibands (secondary Dirac cones) by a magnetic field. An example of the experimental results from the 
magnetoconductance probe of the minigap opening in graphene is shown in Fig.\ \ref{dean}, where the fractal 
pattern is clearly visible. 

\section{Bilayer graphene}

Bilayer graphene consists of two coupled monolayers. This coupling opens a gap in the low energy 
dispersion relation and, in a magnetic field, modifies the LL structure. We consider the bilayer graphene 
with Bernal stacking. A single-particle Hamiltonian (kinetic energy part) of this system 
in a magnetic field is \cite{mccann_06}
\begin{equation}
{\cal H}_{\xi}^{(bi)} = \xi\left( 
\begin{array}{cccc}
\frac{U}2  & v^{}_{\rm F} \pi^{}_{-} & 0 & 0 \\
 v^{}_{\rm F} \pi^{}_{+} & \frac{U}2  &
\xi\gamma^{}_1 & 0 \\ 
 0 &\xi\gamma^{}_1 & -\frac{U}2  &
v^{}_{\rm F}
\pi^{}_{-} \\      
 0 & 0 & v^{}_{\rm F} \pi^{}_{+} & -\frac{U}2 
 \end{array} 
\right),
\label{HAB2}
\end{equation}
where $\xi = \pm 1$ corresponds to two valley ($K$ and $K^{\prime }$), $U$ is the inter-layer bias 
voltage which can be varied for a given system, and $\gamma^{}_1 \approx 0.4 $ eV is the inter-layer 
coupling. The eigenfunctions of the Hamiltonian (\ref{HAB2}) can be expressed in term of the Landau 
functions $\phi^{}_{n,k}$ [Eq.\ (\ref{PhiK})]
\begin{equation}
\Psi^{\rm (bi)}_{n,k}  = 
\left( \begin{array}{c}
 \xi  C^{}_1 \phi^{}_{|n|-1,k} \\
  {\rm i} C^{}_2   \phi ^{}_{|n|,k} \\  
  {\rm i} C^{}_3  \phi^{}_{|n|k} \\
  \xi C^{}_4 \phi ^{}_{|n|+1,k}  
\end{array}  
 \right),
\label{fAB2}
\end{equation} 
where the coefficients, $C^{}_1$, $C^{}_2$, $C^{}_3$, and $C^{}_4$, 
can be found from the following system of equations
\begin{eqnarray}
& &  \varepsilon C^{}_1 =  \xi u  C^{}_1 - \sqrt{n} C^{}_2  
\label{seq1} \\
 & &  \varepsilon C^{}_2 = \xi u  C^{}_2 - \sqrt{n} C^{}_1 + 
 \tilde{\gamma}^{}_1 C^{}_3 
 \label{seq2} \\
 & &  \varepsilon C^{}_3 =  -\xi u  C^{}_3 + \sqrt{n+1} C^{}_4 + 
 \tilde{\gamma}_1 C^{}_2 
 \label{seq3} \\
 & &  \varepsilon C^{}_4 = -\xi u C^{}_4 + \sqrt{n+1} C^{}_3 
 \label{seq4}.
\end{eqnarray}
Here all energies are expressed in units of $\epsilon^{}_B=\hbar
v^{}_{\rm F}/\ell^{}_0$, $\varepsilon$ is the energy of the LL, 
 $u=U/(2\epsilon^{}_B)$, and 
$\tilde{\gamma}^{}_1=\gamma^{}_1/\epsilon^{}_B$. 

The energy spectra of the LLs can be found from \cite{pereira_07}
\begin{equation}
\left[\left(\varepsilon+\xi u \right)^2-2n\right]
\left[\left(\varepsilon-\xi u \right)^2-2(n +1)\right]
= \tilde{\gamma}_1^2 \left[ \varepsilon ^2
- u^2 \right].
\label{eigen1}
\end{equation}
For each value of $n\geq 0$ there are four solutions of the eigenvalue 
equation (\ref{eigen1}), corresponding to four Landau levels in a bilayer 
graphene for a given valley, $\xi=\pm1$. For zero bias voltage, $U = 0$ these four Landau levels are 
\begin{equation}
\epsilon = \pm \sqrt{ 2n+1 + \frac{\tilde{\gamma}^{2}_1 }{2} \pm 
\frac12 \sqrt{(2+ \tilde{\gamma}^{2}_1 )^2 + 8 n \tilde{\gamma}^{2}_1  }   }.
\label{zero}
\end{equation}
In this case each Landau level has two-fold valley degeneracy which is lifted at finite bias voltage $U$.

For $n=0$ there are two special LLs of bilayer graphene. One LL has the energy $\varepsilon = - \xi u$ 
and the wave function of this LL consists of $\phi^{}_{0,k}$ functions only  
\begin{equation}
\Psi^{\rm (bi)}_{0^{}_1,k}=\left( \begin{array}{c}
 \phi^{}_{0,k} \\
 0 \\  
  0 \\
  0  
\end{array}  
 \right).
\label{f00}
\end{equation} 
This LL of bilayer graphene has exactly the same properties as for the $0$-th conventional, 
non-relativistic Landau level. For zero bias voltage $U$, this level has zero energy. 

For small values of $U$ there is another solution of Eq.~(\ref{eigen1}) with $n=0$, which has almost 
zero energy, $\varepsilon\approx 0$. The corresponding LL has the wavefunction 
\begin{equation}
\Psi^{\rm (bi)}_{0^{}_{2},k}  = 
\frac1{\sqrt{\gamma_1^2 + 2 \epsilon_B^2}}
\left( \begin{array}{c}
 \gamma^{}_1 \phi^{}_{1,k} \\
 0 \\  
 \sqrt2 \epsilon^{}_B\phi^{}_{0,k} \\
  0  
\end{array}  
 \right).
\label{f11}
\end{equation} 
The wave function of this LL is the mixture of the $n=0$ and $n=1$ conventional (nonrelativistic) 
Landau functions $\phi^{}_{0,k}$ and $\phi^{}_{1,k}$. This mixing depends on the magnitude of the 
magnetic field. In a small magnetic field, $\epsilon^{}_B \ll \gamma^{}_1$, the 
wavefunction is $(\psi^{}_{1,m}, 0, 0, 0)^T$ and the LL is identical to the $n=1$ non-relativistic 
LL. In a large magnetic field $\epsilon^{}_B \gg \gamma^{}_1$, the LL wavefunction is $(0, 0, 
\psi^{}_{0,m}, 0)^T$ and the bilayer LL has the same properties as the $n=0$ non-relativistic LL. 

Following the same procedure as for the conventional systems and the graphene monolayer, we can find the 
matrix elements of the periodic potential in the basis of LL wave function of bilayer graphene 
\begin{eqnarray}
& & \left\langle n^{\prime }k^{\prime}\right|\cos(q^{}_0 y)\left| n k\right
\rangle = \nonumber \\  
& & \frac{i^{n-n^{\prime}}}{2} C^{}_n C^{}_{n^{\prime}} 
\left\{\delta^{}_{k^{\prime}, k+q^{}_0} 
  +  (-1)^{n-n^{\prime }} \delta^{}_{k^{\prime}, k-q^{}_0} 
  \right\}      \nonumber \\
 & &  \times 
\left[C^{}_{n,1} C^{}_{n^{\prime},1} M^{}_{|n^{\prime}|-1,|n|-1} 
+ C^{}_{n,4} C^{}_{n^{\prime},4} M^{}_{|n^{\prime}|+1,|n|+1}  \right. \nonumber \\
& & + \left. \left( C^{}_{n,2} C^{}_{n^{\prime},2}  + C^{}_{n,3} C^{}_{n^{\prime},3}\right) 
M^{}_{|n^{\prime}|,|n|} 
     \right] 
\end{eqnarray}
and 
\begin{eqnarray}
& & \left\langle n^{\prime}k^{\prime}\right|\cos(q^{}_0 x) \left| n k\right
\rangle = \frac{\delta^{}_{k^{\prime}, k}}{2} C^{}_n C^{}_{n^{\prime}} 
e^{-i q^{}_0 k\ell_0^2} \left[1+(-1)^{n-n^{\prime}}\right]  
\nonumber \\
& &\times 
\left[C^{}_{n,1} C^{}_{n^{\prime},1} M^{}_{|n^{\prime}|-1,|n|-1} 
+ C^{}_{n,4} C^{}_{n^{\prime},4} M^{}_{|n^{\prime}|+1,|n|+1}  \right. \nonumber \\
& & + \left. \left( C^{}_{n,2} C^{}_{n^{\prime},2} + C^{}_{n,3} C^{}_{n^{\prime},3}\right) 
M^{}_{|n^{\prime}|,|n|} \right]. 
\end{eqnarray}

\begin{figure}
\begin{center}\includegraphics[width=7cm]{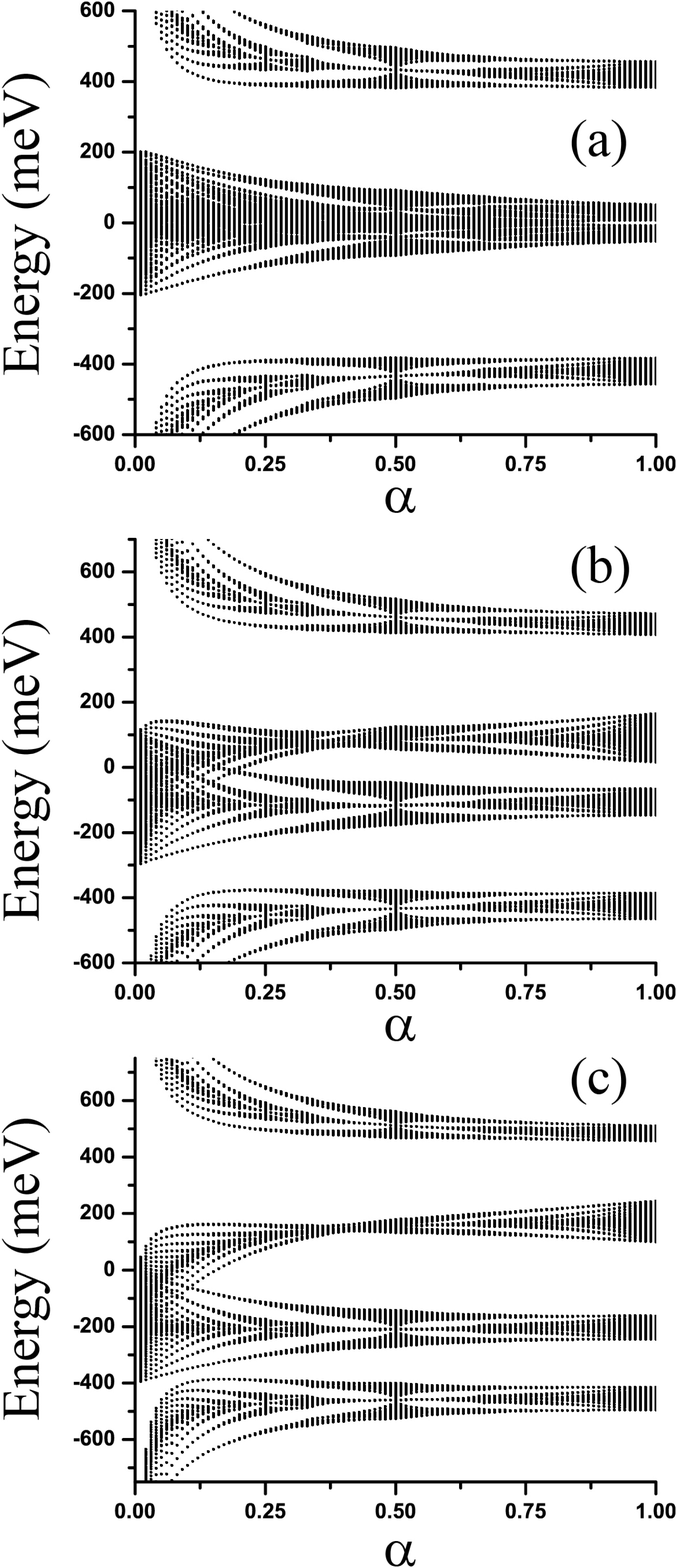}\end{center}
\caption{Single-electron energy spectra of bilayer graphene in a periodic potential and an external
magnetic field. The period of the potential is $a^{}_0 = 20$ nm and its amplitude is $V^{}_0 = 100$
meV. The bias voltage is (a) $U=0$, (b) $U=200$ meV, and (c) $U=400$ meV. The energy
spectra are shown as a function of the parameter $\alpha = \Phi^{}_0 /\Phi$.
}
\label{Fig_bilayer}
\end{figure}

With the known matrix elements of the periodic potential, we can find the energy spectra of bilayer 
graphene in a magnetic field and weak (or intermediate) periodic potential, taking into account many LLs. 
The results are shown in Fig.\ \ref{Fig_bilayer}. For zero bias voltage [Fig.\ \ref{Fig_bilayer}(a)], 
similar to graphene, the inter-Landau level coupling becomes important only for large amplitudes of 
the periodic potential, $V^{}_0> 100 $ meV. This is true except for two degenerate LLs of type (\ref{f00}) 
and (\ref{f11}), for which the inter-level coupling becomes strong even for small amplitudes $V^{}_0$ due 
to the degeneracy of the levels. In this case, the structure of the energy spectrum near zero energy becomes 
complicated due to the mixture of two degenerate butterfly structures. These two butterfly structures are
not identical due to different types of wave functions of the two LLs and correspondingly different
effective periodic potentials. For one LL the effective periodic potential is $V^{}_0L^{}_0(\pi\alpha)$,
while for the other LL, the wave function of which is given by Eq.~(\ref{f11}), the effective strength
of the potential is
\begin{equation}
\frac{V^{}_0}{\gamma_1^2+ 2\epsilon_B^2}
\left(\gamma_1^2L^{}_1(\pi\alpha) + 2\epsilon_B^2L^{}_0(\gamma_1^2L^{}_0(\pi\alpha)\right).
\end{equation} 

At a finite bias voltage [Fig.\ \ref{Fig_bilayer}(b,c)] the degeneracy of two low energy LLs is 
lifted and we can observe two distinctively separated butterfly structures for large values of $\alpha$. 
For small $\alpha $ (large magnetic field), there is a large overlap of the two butterfly structures and 
a strong inter-level mixture is expected. In one of the initially degenerate LLs the flatband condition 
is satisfied for $\alpha \approx 0.35$ [Fig.\ \ref{Fig_bilayer}(c)]. The Hofstadter butterfly in 
bilayer graphene has been studied in \cite{nemec}, where general configuration 
of the bilayer graphene, e.g., continuous displacement between the layers, were introduced. 

\begin{figure}
\begin{center}\includegraphics[width=9cm]{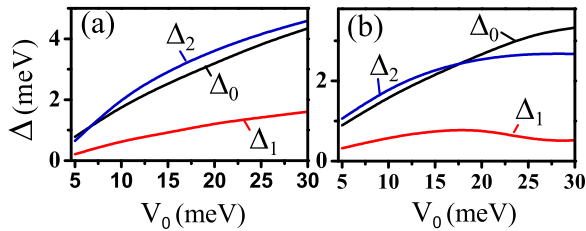}\end{center}
\caption{The band gaps in the $n = 0$, $n = 1$, and $n = 2$
LLs versus the amplitude of the periodic potential,
$V^{}_0$, for interacting systems with half filling of the $n = 0$ Landau
level. The band gap $\Delta ^{}_n$ at LL with index $n$ is defined as the gap between
the corresponding bands of Dirac fermions in a magnetic field
corresponding to $\alpha = 1/2$. The period of the potential is (a)
$a^{}_0 = 20$ nm and (b) $a^{}_0 = 40$ nm.
}
\label{Fig_Flux2}
\end{figure}

\section{Interaction effects}

\subsection{Hartree approximation}

Theoretical analysis of the Hofstadter butterfly problem was mainly restricted to noninteracting 
electron systems. There were only a few papers that reported on the effects of electron-electron 
interactions on the fractal energy spectra \cite{screening_95,vidar,doh_salk,Apalkov_14} The problem 
with inclusion of the electron-electron interaction into the system is related to the requirement that the 
system should have a large size to capture the fractal nature of the spectrum. The Hartree or 
mean-field approaches have been used to estimate the effect of interactions on the electron energy 
spectrum. 

\begin{figure}
\begin{center}\includegraphics[width=8.5cm]{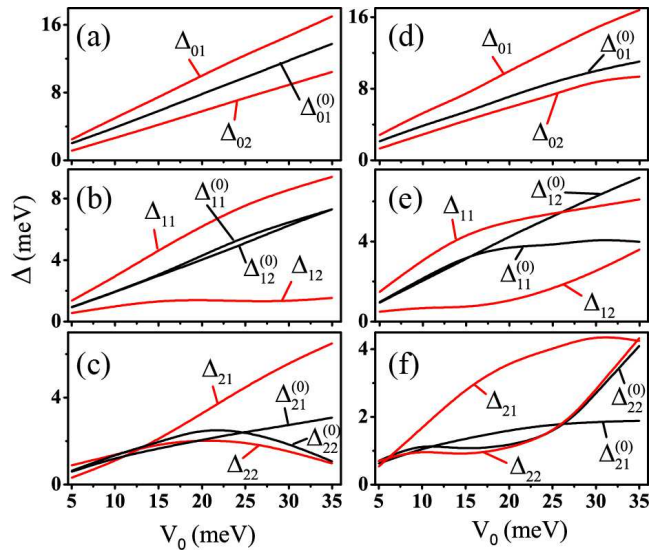}\end{center}
\caption{The band gaps versus $V^{}_0$ for $n = 0$ (a,d), $n = 1$ (b,e),
and $n = 2$ (c,f) LLs. The band gaps are defined as
the gaps between the corresponding bands of Dirac fermions
in a magnetic field for $\alpha = 1/3$. The black lines correspond
to the case of the nonintercting system, while the red lines
correspond to the Dirac fermions with Hartree interaction and
half filling of the $n = 0$ Landau level. The gaps are labeled
as $\Delta^{(0)}_{ni}$ (noninteracting system) and $\Delta^{}_{ni}$ (interacting system),
where $n$ is the LL index and $i = 1$ and $2$ corresponds
to the low-energy and high energy gaps, respectively. The
period of the periodic potential is $a^{}_0 =20$ nm (a,b,c) and $a^{}_0 = 40$ nm
(d,e,f).
}
\label{Fig_L20_40}
\end{figure}

In the Hartree approach the problem is reduced to the single-electron problem in a periodic potential 
and the Hartree potential, produced by the inter-electron interaction with average electron density. 
The Hamiltonian of the system with the Hartree interaction is
\begin{equation}
{\cal H} = {\cal H}^{}_0 (p^{}_x,p^{}_y) + V(x,y) + V^{}_H (x,y),
\label{HgeneralHartree}
\end{equation}
where $V^{}_H(x,y)$ is the Hartree potential, which can be expressed as
\begin{equation}
V^{}_H(x,y) = \int dx^{}_1 dy^{}_1  \frac{e^2}{\kappa |\vec{r} - \vec{r}^{}_1|} n(\vec{r}^{}_1).
\label{VH}
\end{equation}
Here $\kappa$ is the background dielectric constant and 
\begin{equation}
n(\vec{r})= \sum_i^{\prime} |\Psi^{}_i (x,y) |^2,
\label{nR}
\end{equation}
where the prime means that the sum goes over all occupied electron states. The number of occupied 
states is determined by the chemical potential of the system, $\mu$, i.e., only the states with energy 
$E^{}_i$ less than the chemical potential, $E^{}_i < \mu$, are occupied. 
The wave functions $\Psi^{}_i (x,y)$ 
are single particle wave functions of Hamiltonian (\ref{HgeneralHartree}). 

\begin{figure}
\begin{center}\includegraphics[width=9cm]{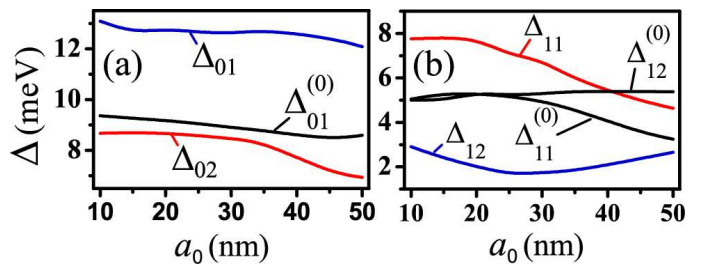}\end{center}
\caption{The band gaps in (a) $n = 0$ and (b) $n = 1$ LLs versus the period $a^{}_0$ of the
periodic potential for non-interacting system and the system with interaction and half filling of 
the $n = 0$ Landau level for $\alpha = 1/3$. The amplitude of the potential is $V^{}_0 = 25$ meV.
}
\label{Fig_flux13vsQ}
\end{figure}

The finite size system (\ref{HgeneralHartree})-(\ref{nR}) can be solved numerically following the
self-consistent procedure. The final solution is the energy spectrum of electron system with the Hartree 
interaction. It is convenient to express the Hartree potential in the reciprocal space. The electron 
density should have the same spatial symmetry as the periodic potential. Then the Fourier transform 
of the electron density 
\begin{equation}
\tilde{n}(\vec{G}) = \frac1{A^{}_0} \int d\vec{r} \, n(\vec{r}) e^{-i\vec{r}\vec{G}},
\label{nG}
\end{equation}
is nonzero only at points of reciprocal lattice, i.e., at points $\vec{G} = \vec{G}^{}_{n^{}_x,n^{}_y} = 
(2\pi/a^{}_0)(n^{}_x,n^{}_y)$, where $n^{}_x$ and $n^{}_y$ are integers. Here $A^{}_0$ in Eq.\ (\ref{nG})
is the area of the sample. Then the Fourier transform of the Hartree potential is also nonzero only at 
points of the reciprocal lattice and is given by 
\begin{equation}
V (\vec{G}) = \frac{2\pi e^2}{\kappa |\vec{G}|} \tilde{n}(\vec{G})  ~~ \mathrm{for}~~ \vec{G} \neq 0,
\end{equation}
and $V(\vec{G} =0) = 0$. In Ref.\ \cite{screening_95,vidar} this approach was used to study the 
interaction effects on Hofstadter butterfly in conventional systems, where strong oscillations of the 
LL bandwidth with chemical potential, i.e., filling of the LL, were reported. 

Following the procedure outlined above, the interaction effects on the band structure of the 
Hofstadter butterfly in graphene were studied in Ref.\ \cite{Apalkov_14}. 
The graphene LLs with indices  $n = 0$, $n = \pm 1$, and $n = \pm 2$ were considered and the gap 
structure for $\alpha = 1/2$ and $\alpha = 1/3$ with interaction and without interaction were analyzed. 
Periodic boundary conditions were applied and the size of the system was $50 a^{}_0 \times 50 a^{}_0$. 
For $\alpha = 1/2$, the system is expected to have two bands separated by a gap. For noninteracting 
system the gap is zero at all LLs. Finite electron-electron interactions open gaps for $\alpha = 1/2$, 
where the magnitude of the gap depends both on the period $a^{}_0$ of the periodic potential and its 
magnitude $V^{}_0$. In Fig.\ \ref{Fig_Flux2} \cite{Apalkov_14} this dependence is shown 
for the case when half of the $n=0$ LL is occupied, i.e., the chemical potential is zero. Strong nonmonotonic 
dependence of the gaps on the LL index is clearly visible in Fig.\ \ref{Fig_Flux2}, and as a function of
the LL index the gap has a minimum for $n=1$. 

The case of $\alpha = 1/3$ has been also studied in Ref.\ \cite{Apalkov_14}. In this case even 
without the interaction, the system has three bands and correspondingly two nonzero gaps in each LL. For 
a non-interacting system the two gaps $i=1,2$ in the LL with index $n$ are labeled as $\Delta ^{(0)}_{n,i}$.
Due to the symmetry the two gaps in the $n=0$ LL are the same, $\Delta ^{(0)}_{01}=\Delta ^{(0)}_{02}$. In 
higher LLs ($n=1$ and 2) the two gaps are different due to the LL mixing introduced by the periodic potential. 
Then the gaps in the same LL are different, e.g., $\Delta ^{(0)}_{11}\neq \Delta ^{(0)}_{12}$.
Interaction modifies the gaps with the general tendency that the lower energy gap is enhanced and the higher 
one is suppressed. For $n=0$ the two gaps are no longer equal, $\Delta ^{}_{01}\neq \Delta^{}_{02}$. 
As a function of the amplitude of the periodic potential the gaps have nonmonotonic dependence with 
local minimum (or maximum) at finite values of $V^{}_0$. The higher energy gap for $n=1$, 
$\Delta^{}_{12}$, is strongly suppressed by the electron-electron interactions. 

The enhancement or suppression of the gaps by the electron-electron interactions depend not only on the 
amplitude of the periodic potential but also on the period of the potential. This dependence is shown in 
Fig.\ \ref{Fig_flux13vsQ} for $\alpha = 1/3$ and amplitude of the potential $V^{}_0 = 25 $ meV. The 
results are shown for the $n=0$ and $n=1$ LLs only. The gaps, both for the system with interactions and 
without interactions, have weak dependence on $a^{}_0$ for small values of the period, $a^{}_0 \lesssim 
25$ nm. For larger values of $a^{}_0$ there is a strong suppression of the low energy gap, $\Delta^{}_{11}$,
in the $n=1$ LL and higher energy gap, $\Delta^{}_{02}$, in the $n=0$ LL. In general, the gaps have monotonic 
dependence on $a^{}_0$, except the higher energy gap, $\Delta^{}_{12}$, in the $n=1$ LL, which has a minimum 
at $a^{}_0 \approx 25$ nm. 

\subsection{Correlation effects: Extreme quantum limit}

In the extreme quantum limit, i.e., in a strong magnetic field and extremely low temperatures, electrons display
the celebrated fractional quantum Hall effect (FQHE), which is an unique manifestation of the collective modes of the 
many-electron system. The effect is driven entirely by the electron correlations resulting in
the so-called incompressible states \cite{FQHE_book,laughlin}. It should be pointed out that the properties of 
incompressible states of Dirac fermions have been established theoretically for monolayer graphene 
\cite{mono_FQHE} and bilayer graphene \cite{bi_FQHE} and the importance of interactions in the extreme quantum 
limit are well known \cite{chapter,interaction}. There are also experimental evidence of the FQHE states 
in graphene \cite{graphene_book,FQHE_expt}. The precise role of FQHE in the fractal butterfly 
spectrum has remained unanswered however. Interestingly, in a recent experiment \cite{geim_14}, the butterfly 
states in the integer quantum Hall regime \cite{vktc} have been already explored. Understanding the effects of 
electron correlations on the Hofstadter butterfly is therefore a pressing issue. In Ref.~\cite{butterfly_FQHE},
the authors have recently developed the magnetic translation algebra \cite{mag_translation,haldane_85,read} of 
the FQHE states, in particular for the primary filling factor $\nu=\frac13$ for Hofstadter butterflies in 
graphene \cite{butterfly_FQHE}. They considered a system of electrons in a periodic rectangular geometry that 
was very useful earlier in studying the properties of the FQHE in the absence of a periodic potential
\cite{prg}. The work in Ref.~\cite{butterfly_FQHE} has unveiled a profound effect of the FQHE states on the 
butterfly spectrum resulting in a transition from the incompressible FQHE gap to the gap due to the periodic 
potential alone, as a function of the periodic potential strength. There are also crossing of the ground state 
and low-lying excited states depending on the number of flux quanta per unit cell, that are absent when the 
periodic potential is turned off. 

The magnetic translation analysis was employed to study the effect of a periodic potential on the FQHE in 
graphene for the primary filling factor $\nu=1/3$. For $\alpha=1/2$ and $\alpha=1/3$, increasing the periodic 
potential strength $V^{}_0$ resulted in a closure of the FQHE gap and the appearance of gaps due to the periodic 
potential \cite{butterfly_FQHE}. It was also found that for $\alpha=1/2$ this results in a change of the 
ground state and consequently in the change of the ground state momentum. For $\alpha=1/3$, despite the 
observation of the crossing between the low-lying energy levels, the ground state does not change with an 
increase of $V^{}_0$ and is always characterized by zero momentum. The difference between these two $\alpha$ s 
is a result of the origin of the gaps for the energy levels. For $\alpha=1/2$ the emergent gaps are due to the 
electron-electron interaction only, whereas for $\alpha=1/3$ these are both due to the non-interacting Hofstadter
butterfly pattern and the electron-electron interaction.

\section{concluding Remarks}

It has been a while since the beautiful theoretical idea of the Hofstadter butterfly which encompasses the fractal
geometry and electron dynamics in a magnetic field and periodic potentials was proposed and its eventual confirmation 
in real physical systems in recent years. The unusual properties of graphene described here actually helped in finding 
the exotic butterflies with their fractal pattern. In achieving this feat the experimentalists made tremendous progress 
in understanding and controlling the properties of graphene under various constraints, and creating the Moir\'e pattern 
by finding the right substances. Discovery of the fractal butterfly in graphene has opened up new directions of research, 
both in materials research, and in fundamental studies of two-dimensional electrons. Future experiments will undoubtedly 
be in the limit of strong electron correlations, thereby opening up the fertile field of many-body effects in Dirac 
materials. 

The work has been supported by the Canada Research Chairs Program of the Government of Canada. We wish to thank 
Philip Kim and Cory Dean for sending us a copy of Fig. 5. The work of V.M.A. has been supported by the NSF grant 
ECCS-1308473.

\end{document}